\documentclass[aps,prc]{revtex4}
\usepackage{epsfig}
\begin{document}

\title{
 The reaction $^2$H(p,pp)n in three kinematical configurations 
at E$_p$ = 16 MeV}

\author
{C. D\"uweke, R. Emmerich, A. Imig, 
J. Ley, G. Tenckhoff, H. Paetz  gen. Schieck\footnote{Corresponding author}}
\affiliation{Institut f\"{u}r Kernphysik, Universit\"{a}t zu K\"{o}ln, 
Z\"{u}lpicher Stra{\ss}e 77, D-50937 K\"{o}ln, Germany} 
\author
{J. Golak, H. Wita{\l}a}
\affiliation{Institute of Physics, Jagellonian University, 
Reymonta 4, PL-30059 Cracow, Poland}
\author{E. Epelbaum}
\affiliation{Jefferson Laboratory, Theory Division, Newport News, VA 23606, USA}
\author{W. Gl\"{o}ckle}
\affiliation{Institut f\"{u}r Theoretische Physik, Ruhr-Universit\"{a}t Bochum 
Universit\"{a}tsstra{\ss}e 150, D-44780 Bochum, Germany}
\author{A. Nogga}
\affiliation{Institute for Nuclear Theory, University of Washington, 
Seattle, WA 98195-1550, USA}

\begin{center}
{\bf \today}
\end{center}

\vspace*{1cm}

\begin{abstract}
     
We measured the cross sections of the $^2$H(p,pp)n breakup reaction 
at E$_p$=16 MeV in three kinematical configurations: the np final-state 
interaction (FSI), the co-planar star (CST), and an 
intermediate-star (IST) geometry. 
The cross sections are compared with theoretical predictions 
 based on the CD Bonn potential alone and combined with the updated 
2$\pi$-exchange Tucson-Melbourne three-nucleon force (TM99'), 
calculated without inclusion of the Coulomb interaction.  
The resulting excellent agreement between data and pure CD Bonn predictions 
 in the FSI testifies to the smallness of three-nucleon force (3NF) 
effects as well as the insignificance of the Coulomb force 
for this particular configuration and energy. The CST also 
agrees well whereas the IST results show small deviations between 
measurements and theory seen before in the pd breakup space-star geometries 
which point to possible Coulomb effects. An additional comparison with EFT predictions (without 3NF) up to order N$^3$LO shows excellent agreement in the FSI case and a rather similar agreement as for CD Bonn in the CST and IST situations.  

\end{abstract}

\maketitle

\section{Introduction}

The three-nucleon system has for a long time been the testing ground for 
nucleon-nucleon forces as they act in the lightest systems composed of 
nucleons. 
State of the art calculations -- using Faddeev techniques in 
momentum space -- of the three-nucleon  scattering  allow the comparison 
 of numerically exact theoretical predictions 
with a wide range of observables such as elastic or breakup cross sections 
and polarization data.
Though the overall agreement between theory and experiment is rather 
good there remain certain discrepancies\cite{glo96}. The most prominent 
one is the drastic underestimation by all modern NN potentials of the vector 
analyzing power $A_y$ in low-energy (i.e. at energies below $\approx$ 25 MeV)
 nucleon-deuteron (Nd) elastic 
scattering\cite{glo96}. 
Both in the proton-deuteron (pd) and neutron-deuteron (nd) reactions theory lies 
about 30\% below the data in the angular region of 
the analyzing power maximum\cite{wit94,kij95,kij96}. Also the 
minima of the Nd elastic scattering angular distributions for laboratory energies 
starting from about $60$ MeV are underestimated by pure two-nucleon (2N) 
force Faddeev calculations\cite{wit98,nem98}. 

For the breakup 
process the low-energy nd space-star configuration in which all three 
nucleons are emerging in the c.m. system with equal momenta in a plane 
perpendicular to the 
incoming beam direction presents a puzzle. Modern NN 
forces give too small nd space-star cross sections at low 
energies\cite{ste89,str89,geb93,set96,zho01}. On the other hand, in this energy region the pd breakup data are systematically below the theoretical nd predictions thus pointing to the importance of Coulomb effects \cite{rau91,gro96,pat96,ish03}.

With a recent measurement of the nd breakup in 
an nn quasi-free scattering configuration \cite{vwi02}, together with 
earlier unpublished Bochum data \cite{lub92}, a new discepancy may show up. 
It is remarkable that recent predictions using the chiral perturbation 
theory approach do not remedy the situation \cite{epe02}.  

These discrepancies point to the necessity of 
including new ingredients in addition to the ``standard'' 
two-body input into 
the 3N continuum calculations such as e.g. three-body forces. However,
 present-day 3NF's show only small effects on the low-energy elastic 
scattering vector analyzing power \cite{tor98}. On the other hand, adding the $2\pi$-exchange 
Tucson-Melbourne (TM) 3NF \cite{coo79,coo81}, adjusted to reproduce 
the experimental triton binding energy \cite{nog97}, fills in the cross section 
minima at higher energies\cite{wit98}. 
 
Whereas good systematic studies exist for the symmetric space star 
(SST), the IST and CST configurations have not yet been explored in the 
pd breakup at low energies. 
Therefore, kinematically complete cross section measurements in these 
configurations were performed. In addition, because in former investigations 
at other energies \cite{glo96,rau91,gro96,pae01} there has been very good 
agreement between $^2$H(n,nn)$^1$H and $^2$H(p,pp)n data themselves as well as 
when compared with 
results from modern Faddeev calculations with realistic NN potentials for 
 FSI geometry we  measured it simultaneously.

In section II we shortly describe the underlying theory.
In section III and IV the details of the experiment 
and of the data analysis are given. The resulting data are 
compared to the theory in section V. We summarize in section VI.  

\section{Theory}

The theoretical predictions presented in this work are based on solutions
of 3N Faddeev equations in momentum space using a realistic NN interaction alone and combined with a 3N force.

The long-range Coulomb force acting between two protons is totally neglected.
In the following  we give a short outlook of the underlying formalism 
and of the numerical performance. For more details we refer to \cite{glo96,wit88,glo83,hue93,hue97}, and references therein.

The break-up operator $\tilde T$ sums up all multiple scattering contributions,
 which originate from the  interactions of three nucleons
via 2N and 3N forces, through the integral equation \cite{hue97}
\begin{eqnarray}
\label{e2}
   \tilde T & = & t P +(1+t G_0){V_4^{(1)}}(1+P) + 
tPG_0\tilde T+(1+t G_0){V_4^{(1)}}(1+P)G_0\tilde T
\end{eqnarray} 
Here $t$ is the NN $t$-matrix , $G_0$ is the free 3N propagator, and $P$ 
is the sum of a cyclical and anticyclical permutations of three objects.
The 3NF is split into three parts 
\begin{eqnarray}
\label{e3}
   V_4=\sum_{i=1}^{3} V_4^{(i)}
\end{eqnarray}
where each one is symmetrical under the exchange of two particles. For 
instance, in the case of $\pi-\pi$ exchange 3NF\cite{coo79},\cite{coo81},  
such a decomposition corresponds to the three possible choices of 
the nucleon undergoing off-shell $\pi-N$ scattering.

The transition operator for the breakup process  $U_0$ can be expressed 
in terms of a $\tilde T$ operator as  \cite{glo96}, \cite{hue97}
\begin{eqnarray}
\label{e1}
   U_0 & = & ( 1 + P)\tilde T .
\end{eqnarray}

We solved Eq.(\ref{e2}) in a partial-wave projected momentum-space basis at 
the nucleon laboratory 
energy $E_{N}^{lab}=16$ MeV using the modern CD Bonn \cite{mac96} interaction. 
This 
interaction reproduces the NN data set with a high 
accuracy as measured by a $\chi^2$/datum very close to 1. 
 In the calculations all partial-wave states with total angular momenta in the 
two-nucleon subsystem up to $j_{max}=3$ were taken into account. 
We checked that this restriction is sufficient at the energy of the 
present study. In the calculations charge independence breaking (CIB) of the NN 
interaction
in the state $^{1}S_{0}$ was taken into account by including the total 3N isospin 
T=3/2 contribution in this particular partial wave and using pp and np CD Bonn 
$^1S_0$ interactions~\cite{wit81}.

To check the magnitude of 3NF effects we combined the CD Bonn potential with the 
updated version of the TM 3NF (TM99')~\cite{fri99,coo01}. In this 
calculation we neglected the total isospin T=3/2 contribution in the $^1S_0$ 
state. Such a restriction to the np force 
for the $^1S_0$ state does not have a significant effect on the CST and IST cross 
sections but is decisive for the np FSI cross sections.   

In recent years significant progress has been achieved towards a deeper understanding of the 
low-energy nuclear dynamics within the chiral Effective Field Theory (EFT) framework. This approach 
is based upon the approximate and spontaneously broken chiral symmetry of Quantum Chromodynamics. 
As shown by Weinberg \cite{Weinb91} in the early 1990's, nuclear forces can be systematically derived 
within the chiral expansion. The smallness parameter is given by the ratio
of a generic low-momentum scale corresponding to external nucleon momenta or the pion mass 
to a hard scale associated with  the $\rho$-mass. Very promising results
have already been achieved for few-nucleon systems in this framework, see e.g.~\cite{epe02,epe01,EM03,Ep04}. 
For a recent review on this and related topics the reader is referred to \cite{BvK}. While
in  our previous studies only np forces have been worked out, in \cite{Ep04,epe03_2} also isospin
breaking effects have been included and thus nn, np and pp forces are now available. 
Further, we have used a novel regularization scheme for pion-loop
integrals proposed in \cite{epe03}, which allows for a better separation between the long- and short-distance 
contributions compared to dimensional regularization.
In the present work, we apply for the first time pp and np forces at 
next-to-leading order (NLO), next-to-next-to-leading order (N$^2$LO) and 
next-to-next-to-next-to-leading order (N$^3$LO) in the chiral expansion  from \cite{Ep04,epe03_2}
to study 3N scattering. We remind the reader that chiral 3N forces 
should be taken into account at N$^2$LO and N$^3$LO. In fact, one important advantage of the 
EFT framework compared to the conventional approach is attributed to the fact, that it allows 
to include three- and more-nucleon forces \emph{consistent} with the NN interactions. 
The inclusion of the chiral 3N forces at N$^2$LO and, especially, at N$^3$LO requires, 
however, additional quite extended work and will be deferred to upcoming publications. 
It should, therefore, be understood, that the EFT results at N$^2$LO and N$^3$LO 
presented in this work are incomplete. We, nevertheless, think that it is 
justified to present already the EFT predictions based solely on the NN force
for three consecutive orders in the chiral expansion. We also remind the reader that, 
at least for conventional interactions, the effects of the 3N force for the 
observables considered are rather modest.
 
\section{Experiment}
The measurements were performed at
the Cologne FN tandem Van de Graaff accelerator facility. The
unpolarized protons were produced by a sputter ion source and accelerated 
to a laboratory energy of 16.0
MeV. We used typical beam currents of 150 nA focussed into a beam
spot of 2 mm diameter inside an Ortec 2800
scattering chamber. Behind the scattering chamber the beamline was
terminated by a Faraday cup.

\begin{figure}[htb]
\epsfig{file=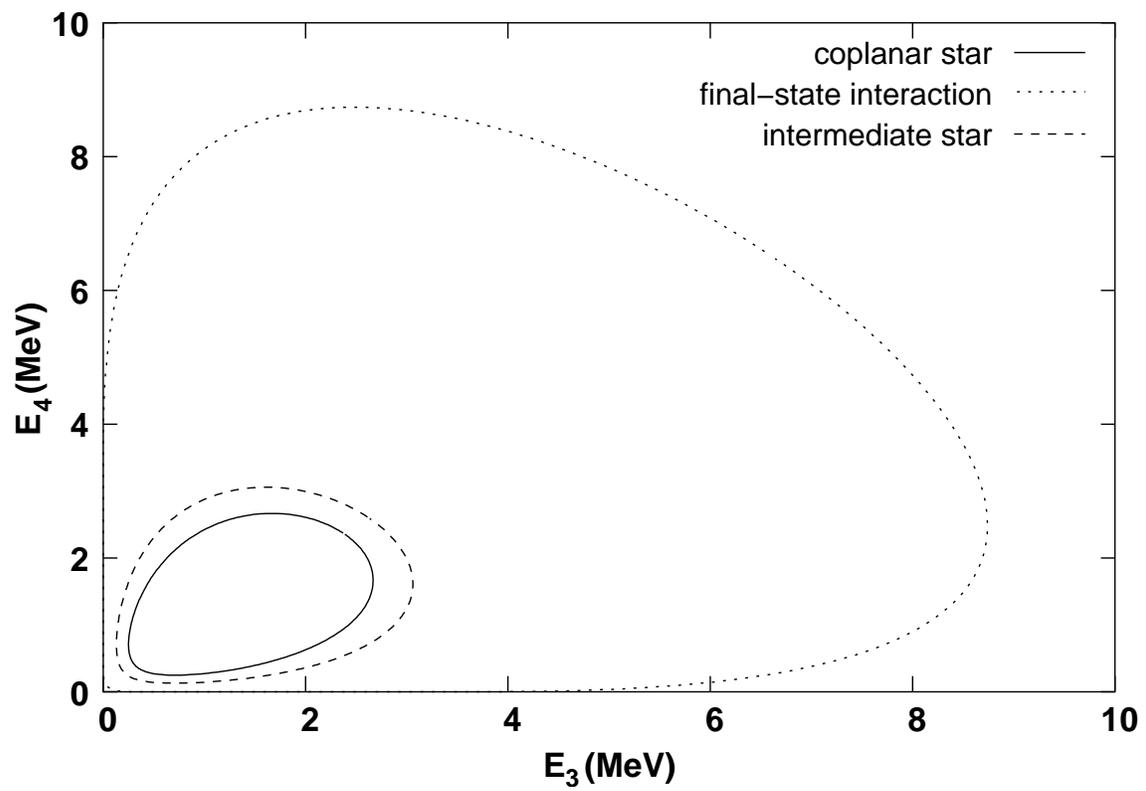,width=15cm}

\vspace{2cm}

\caption{\label{kinloc}Calculated kinematic loci for the three situations 
inverstigated.}
\end{figure}

The target foils used in our measurement consisted of solid deuterated
polyethylene $(CD_{2})_{n}$ with a  $(CD_{2})$ thickness of $600 - 800
\frac{\mu g}{cm^{2}}$ and a carbon backing layer of  $\sim 30 - 40
\frac{\mu g}{cm^{2}}$ on each side and were mounted in a triple target holder. 
Rotation of the targets with about 700 rpm provided for much increased target 
lifetime. For the detection of protons of
the pd breakup reaction and the elastic and inelastic proton
scattering from carbon we used  
2000 $\mu$m thick room-temperature 
silicon surface barrier detectors with an energy resolution  of better than
$\sim$ 50keV (FWHM). The detectors were positioned with an accuracy of $\pm$
0.1\symbol{23} and $\pm$1 mm in the scattering chamber in a symmetrical 
arrangement for 2$\times$3 coincidences. 
Table \ref{winkel-tab} shows
the scattering angles for all coincidence detectors. 
Fig. \ref{kinloc} shows the calculated kinematical loci for the three 
experimental situations.

\begin{figure}[h!]
\epsfig{file=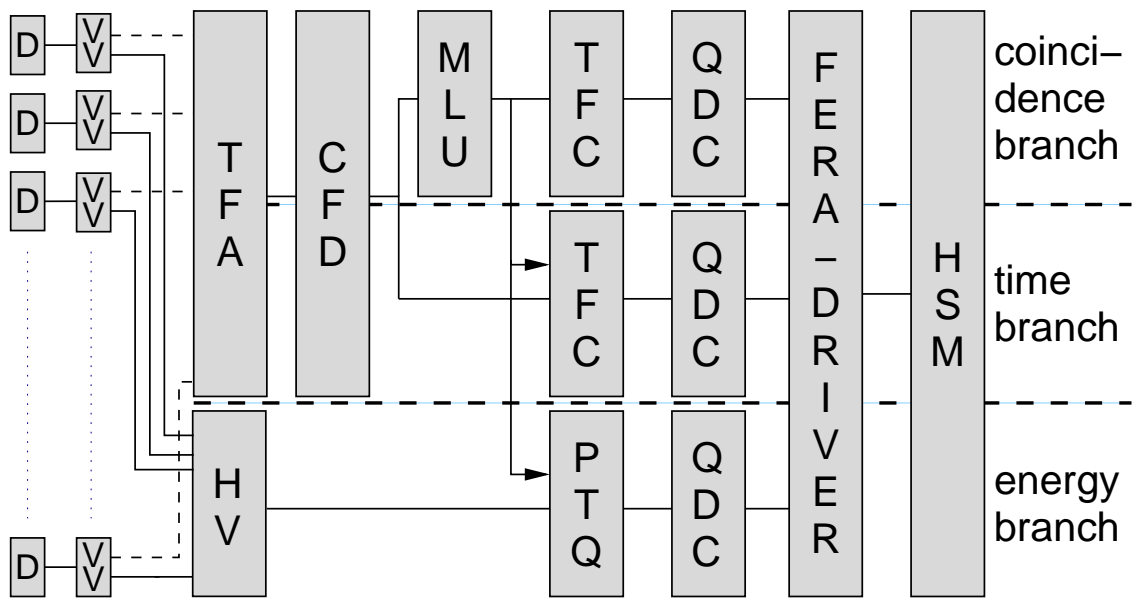,width=15cm}

\vspace{2cm}

\caption{\label{analyzer}Block diagram of the newly-developed electronics 
used for data analysis.}
\end{figure}

Two additional detectors at a laboratory angle of $\Theta$ = 30$\symbol{23}$, 
$\Phi$ = 0$\symbol{23}$ and 180$\symbol{23}$ served as monitors for the absolute
 normalization of
the breakup cross section. Each detector had a total counting rate of less
 than 8kHz and therefore no significant dead-time or
pile-up effects occurred. The signals from all detectors were
processed simultaneously by our new coincidence
electronics and recorded in list mode on magnetic tape. A block diagram of the 
electronics
is shown in Fig. \ref{analyzer}.

The main features of this new setup are the use of QDCs (charge-to-digital 
converters) and an MLU (memory look-up unit) wich allows free programming of 
arbitrary coincidences between up to 16 detectors in very short time.

Each coincidence
event consisted of a logical status word containing the kinematical
configuration, the energies and the time-of-flight differences of the
particles detected. In order to control the experiment on-line, the 
software package PAW (Physics Analyzing Workstation) \cite{cer95} was
used. The coincidence events were 
checked by incrementing and displaying two-dimensional energy spectra,
background-corrected S-curve projections, and distance-parameter spectra. 

\section{Data Analysis}

The coincidence events were stored as triplets ($E_{3},E_{4},\Delta t$)
event by event on magnetic tape together with a status word.
At vanishing angular- and energy-resolutions 
in a kinematically complete measurement the breakup events appear
on ideal point-geometry  kinematical loci in the $E_{3}, E_{4}$ plane. The
measured events scatter around these ideal (point-geometry) kinematical
loci mainly due to finite detector apertures.

For the projection of the breakup events onto 
the kinematical curve (S curve) we assumed a two-dimensional Gaussian
distribution of the data around this curve. 
Therefore the correct projection of a true ($E_{3},E_{4}$) event was
done by assigning it the proper location on the point-geometry
kinematical curve by using ``reference matrices''. For
every event in the calibrated  $E_{3}$, $E_{4}$ plane they contain the
numerically calculated information about the shortest distance to the
S curve. 

\subsection{Time-difference spectra and background subtraction}

\begin{figure}[htb]
\begin{center}
\epsfig{file=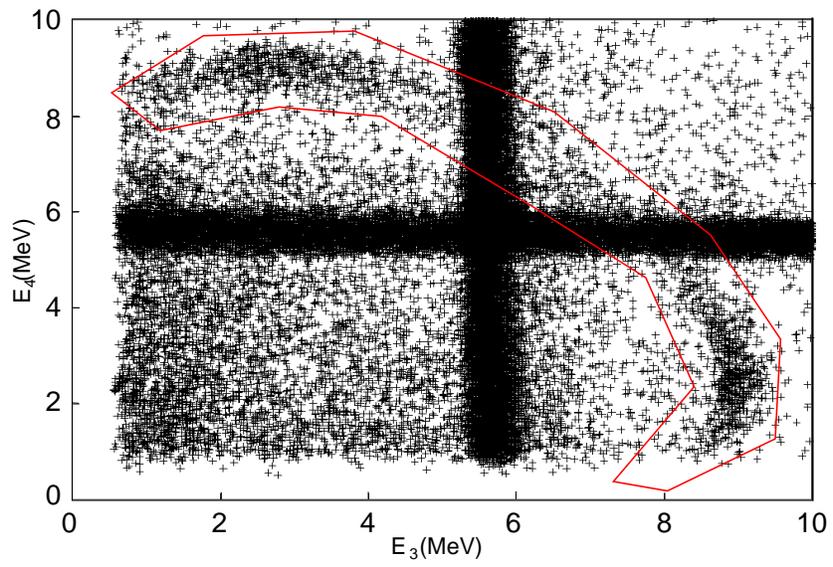,width=11cm}

\vspace{5mm}

\caption{\label{Energiematrix} Calibrated $E_{3} - E_{4}$ coincidence
  matrix for the proton lab. angle pair 51.5\symbol{23}/51.5\symbol{23}.}
\end{center}
\end{figure}

\begin{figure}[htb]
\begin{center}
\epsfig{file=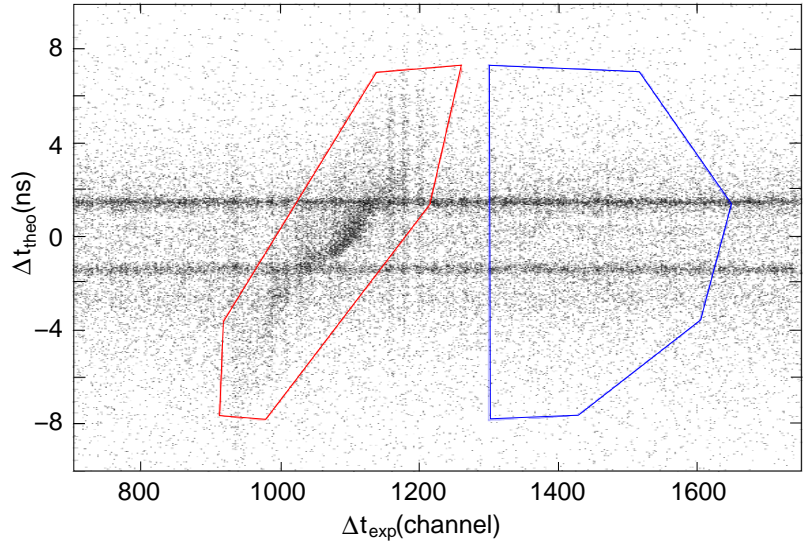,width=11cm}

\vspace{5mm}

\caption{\label{Zeitmatrix}Typical time-difference matrix showing the area of 
the (true+random) events C(tr+r) as well as the area of the randomly distributed 
events C(r). The event density is plotted as a function of time-of-flight 
differences as measured directly, in arbitrary units, and those calculated 
from measured energies and distances, assuming particle masses to be the 
nucleon masses. }
\end{center}
\end{figure}

Fig. \ref{Energiematrix} shows a typical E$_3$, E$_4$ event matrix with 
true coincident events along the kinematical locus and background from 
accidental coincidences. This background consisted mainly of
random coincidences due to elastic scattering and reactions with the
target nuclei $^{2}H$ and $^{12}C$ and truly random coincidences from 
various sources. The background-correction
procedure is based on methods similar to those described in
\cite{pat96,gol83,prz99} and is therefore summarized briefly. 
First 
time-of-flight differences are calculated from the known distances of 
the detectors to the target and under the assumption that the particle
 masses were those of the detected nucleons of the pd breakup. Thus a 
linear relation between the calculated and measured time-of-flight 
differences is expected. When building two-dimensional time-difference 
matrices by sorting the events according to their calculated 
time-of-flight differences and the directly measured  time-of-flight 
differences (Fig. \ref{Zeitmatrix}) the true breakup events therefore 
produce a straight ridge above the uniformly distributed random coincidence 
events. The smaller enclosed region on the left in Fig. \ref{Zeitmatrix} 
encompasses this peak but also contains a contribution from random 
coincidences which has to be subtracted. The larger polygon marked C(r) 
contains only random events. It has the area of the ``true + random'' 
polygon C(tr+r) suitably enlarged by a factor V in the $\Delta$t$_{exp}$ 
direction (with V as 
large as possible). The error from the random-background subtraction 
enters only with this reduction factor V. 

\subsection{Distance-parameter spectra}
In order to subtract the accidental background properly all events in the
 two time-difference windows at a given distance from the kinematical
 locus are summed up along this curve. The ``true'' distance-parameter 
spectrum 
is obtained by subtracting the 
``accidental'' from the ``true+accidental'' spectrum. Only events between 
the dotted lines are then used for projections onto the S curve.

\subsection{S-curve projection}

Thus, after application of our projection procedure to both regions C(tr+r) 
and
C(r) the number of true pd breakup events is given by
\begin{equation}
N_{tr}( \Delta S_{\mu} ) \; = \; N_{tr + r} ( \Delta S_{\mu} ) \, - \,
\frac{1}{V} N_{r} ( \Delta S_{\mu} )
\end{equation}

\noindent with a statistical error of 

\begin{equation}
\Delta_{S} N_{tr}( \Delta S_{\mu} ) \; = \; \sqrt{ N_{tr + r} ( \Delta S_{\mu}
) \, + \, \frac{1}{V^{2}} N_{r} ( \Delta S_{\mu} )}
\end{equation}

\noindent where $\Delta S_{\mu}$ refers to discrete bins with this width on the S 
curve.

As discussed in \cite{prz99} the choice of the binning width $\Delta S_{\mu}$ 
is -- within certain 
limits -- somewhat arbitrary but should be governed by the criterion that narrow 
structures such as the FSI peaks should not be distorted. Here a bin width 
of 200 keV was chosen.\\ 

\begin{figure}[htb]

\begin{center}
\epsfig{file=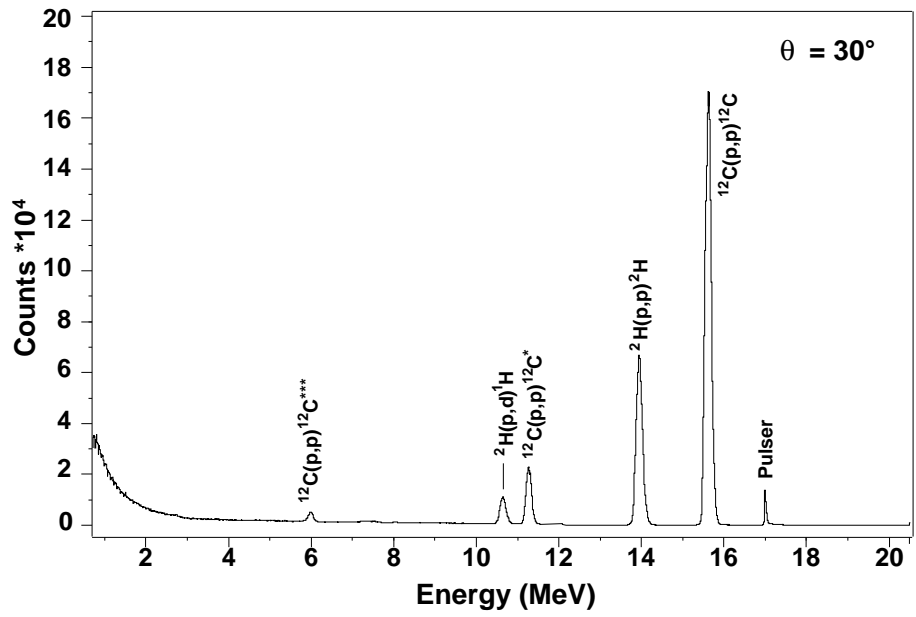,width=12cm}

\vspace{5mm}

\caption{\label{eich} Typical spectrum of a monitor detector at 
$\theta_{lab}=30^\circ$ as used for the cross section calibration.}
\end{center}

\end{figure}

The yields of our S-curve spectra were normalized using the relation
\begin{equation}
\frac{d^{3} \sigma}{d \Omega_{3} d \Omega_{4} d S} \; = \;
\frac{N_{34}}{\Delta \Omega_{3} \cdot \Delta \Omega_{4} \cdot \Delta S_{\mu}}
\frac{\Delta \Omega_{mon}}{N_{mon}} \left( \frac{d \sigma}{d \Omega}
\right)_{mon} 
\end{equation}
where $ ( \frac{d \sigma}{d \Omega})_{mon}$ denotes the differential cross
section of the monitor reaction, $\Delta \Omega_{mon}$ is the solid
angle of the monitor detector and $N_{mon}$ is the background and
dead-time corrected monitor peak intensity. $N_{34}$ is the true (i.e. 
background and dead-time corrected) pd breakup intensity over an
interval $\Delta S_{\mu}$ of the arc length S. $\Delta \Omega_{3}$ and
$\Delta \Omega_{4}$ are the solid angles of the coincidence
detectors. We used the monitor reaction $^{2}$H(p,p)$^{2}$H for which
the elastic scattering cross section was published \cite{cah71,sag94}. 
A typical monitor spectrum used for calibration is shown in Fig. \ref{eich}. 
Explicitly we used the elastic scattering cross
section value at 16.0 MeV and $\Theta$=30$\symbol{23}$ of
\begin{equation}
\left( \frac{ d \sigma }{ d \Omega } \right)_{mon} \; = \; (83.219 \pm 1.248) 
\; \; \; mb/sr
\end{equation}
The errors of the results for the breakup cross section are of different 
origins. The main 
systematic error is due to  the
normalization cross section and the errors of the solid angles. Additionally
we have another systematic error from the projection procedure due to the
distribution of the breakup events around the kinematical point-geometry
loci and uncertainties of their assignment to the correct location there. 
Therefore we had to choose a maximum distance from the S curve within
which the true breakup events were expected. 
We took a distance of typically 0.8 MeV.

The possible loss of breakup events
due to these cuts was checked by varying the 
maximum distance from the S curve.  For the true breakup events this
loss is less than  1\%.  Statistical errors were assumed to originate from
the absolute breakup yield and the background subtraction. These errors 
depend
on the bin width chosen; in our case with $\Delta S_{\mu}$ = 400 keV (CST, IST) and 200 keV (FSI) typical relative errors between 1 and 3\% were assigned.
A study of the effects of averaging over the finite target-detector geometry showed that only in the FSI situation and only at the FSI peaks proper the cross section is lowered by at most 1.2\%. Therefore, here the comparison is made with point-geometry calculations only.

\section{Results and Discussion}

We compare our data with theoretical predictions in two ways; one is
the ``classical'' Faddeev calculation using modern precision NN potentials, with and without a three-nucleon force, the other a new effective-field theory (EFT) calculation up to order N$^3$LO.

\subsection{Comparison to Faddeev calculations with CD Bonn and TM99'}

Since the results of a number of former systematic studies had shown that in the present breakup
 configurations the use of different modern precision potentials 
 leads to very small cross-section changes we compare our results 
to Faddeev calculations with the CD Bonn potential only. For the 3NF the modified Tucson-Melbourne TM99' force is used. In the following three figures also the percentage deviation $\Delta$ of the cross sections caused by the TM99' 3NF is shown.

\begin{equation}
\Delta = { {\sigma(NN+3NF) - \sigma(NN) } \over {\sigma(NN) } } * 100 \%
\label{eqdelta}
\end{equation}

\begin{figure}[htb]

\begin{center}
\epsfig{file=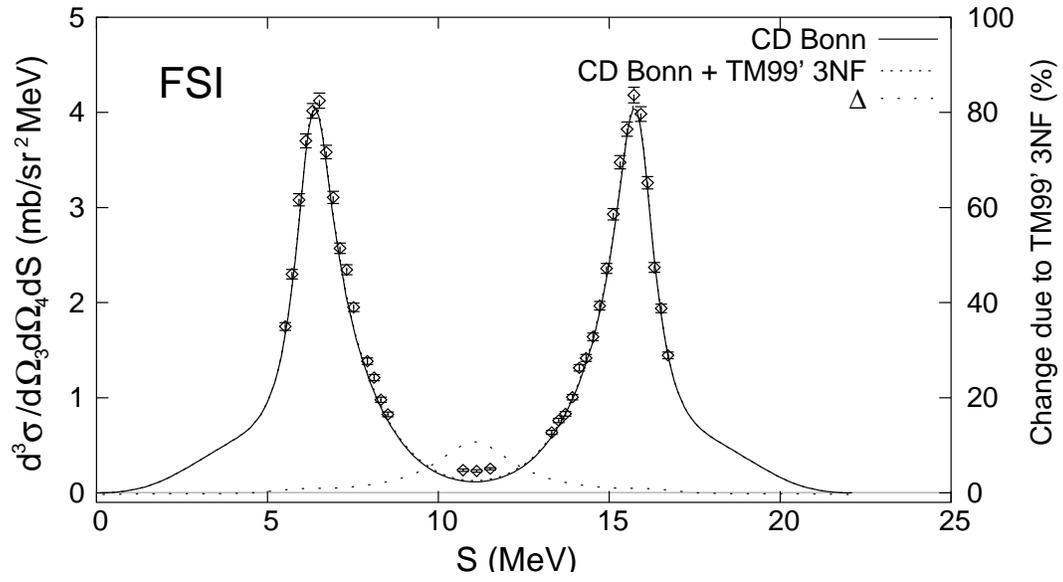,width=14cm}

\vspace{5mm}

\caption{\label{fsi} Results of the breakup cross section
  $\frac{d^{3} \sigma}{ d \Omega_{3} d \Omega_{4} d S}$ 
(see table I) in the FSI situation. Experimental data are
  compared with the results of Faddeev calculations using the 
Bonn CD potential 
with and without the TM99' 3NF. The small effects of the 
latter are depicted also as a percentage deviation $\Delta$.}
\end{center}

\end{figure}

\begin{figure}[htb]
\begin{center}

\epsfig{file=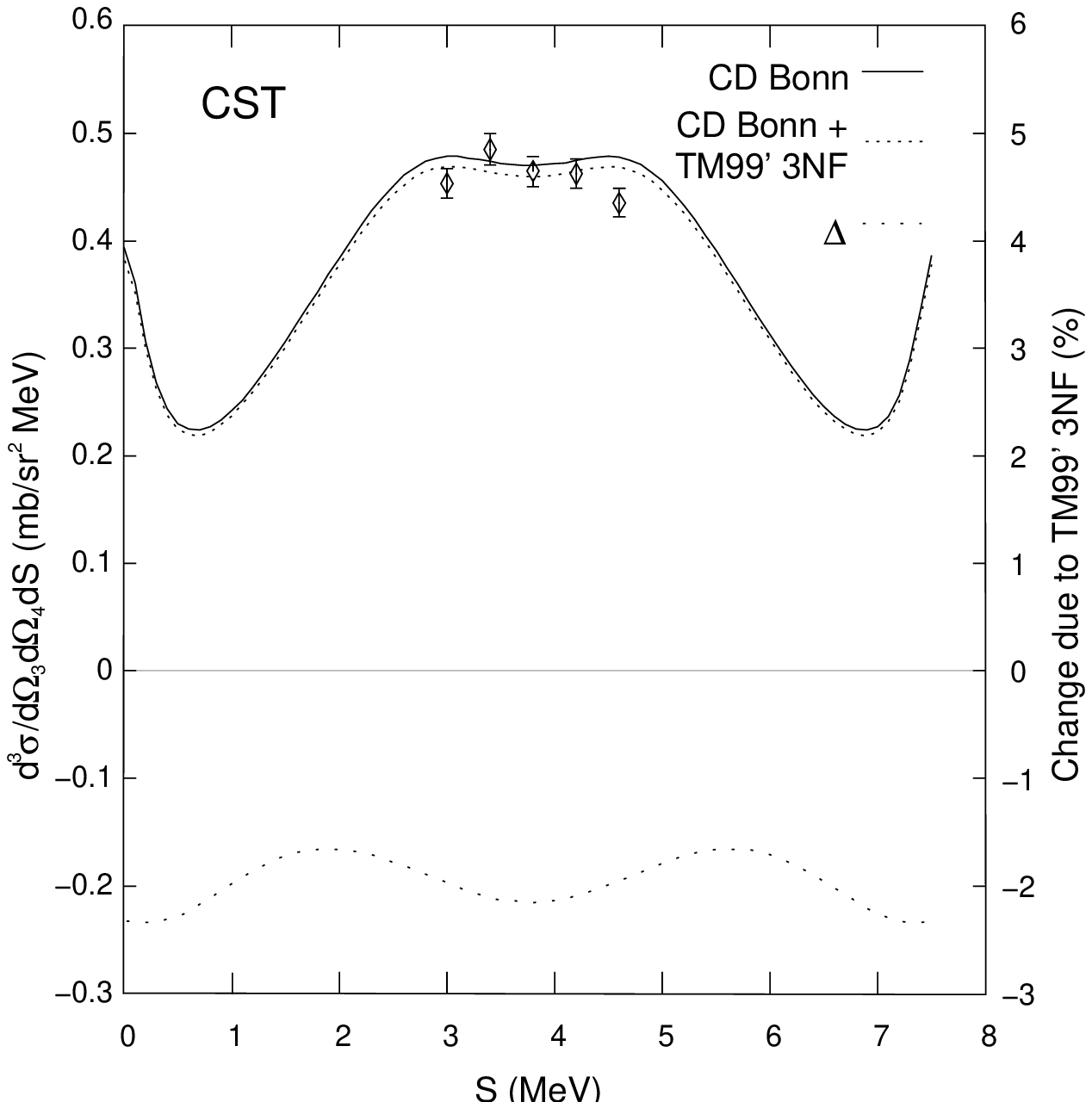,width=14cm}

\vspace{5mm}

\caption{\label{cst} Results of the breakup cross section
  $\frac{d^{3} \sigma}{ d \Omega_{3} d \Omega_{4} d S}$ 
(see table I) in the CST  situation, similar to Fig. \ref{fsi}.}
\end{center}
\end{figure}

\begin{figure}[htb]
\begin{center}

\epsfig{file=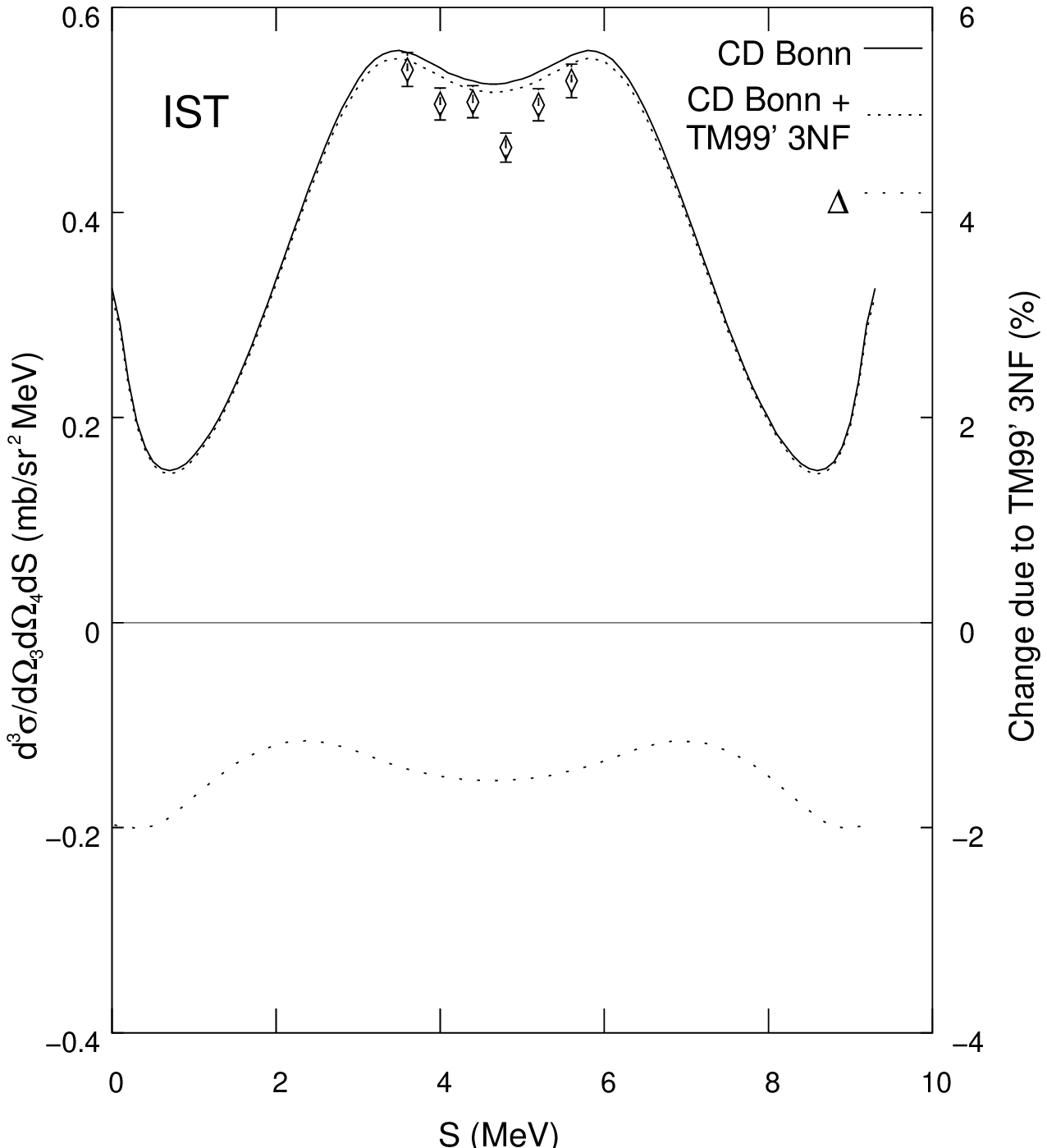,width=14cm}

\vspace{5mm}

\caption{\label{ist} Results of the breakup cross section
  $\frac{d^{3} \sigma}{ d \Omega_{3} d \Omega_{4} d S}$ 
(see table I) in the IST  situation, similar to Fig. \ref{fsi}.}
\end{center}
\end{figure}

In Figs. \ref{cst}, \ref{ist}, and \ref{fsi} we present
our pd breakup cross-section data for the CST, the IST, and the FSI  situations and compare them to point-geometry theoretical predictions. The agreement between experiment and theory is excellent for the FSI and CST situations which would suggest that at these FSI and CST situations Coulomb-force effects do not play a significant role. In the IST there is a slight overestimate of the theory over the data which may be caused by Coulomb effects not contained in the theory. 

It is seen that the effects of the TM99' 3NF are very small for most regions of the S-curve.
 For the FSI peaks and CST and IST geometries the changes of the cross section caused 
by the TM99' 3NF are smaller than $\approx 2 \%$. Only in the region of the S-curve 
between two np FSI peaks effects of up to $\approx 10 \%$ are seen and the TM99'
 brings the theory closer to data.

Finite-geometry corrections in the CST and IST configurations proved to be
insignificant. In the FSI situation there is a small (1.2\%) averaging effect in the relatively
narrow structures of the FSI peaks proper which does not affect significantly the excellent
agreement between the data and the predictions.

\subsection{Comparison to EFT predictions}
It is interesting that the EFT calculations in all orders also show excellent agreement 
with the data in the FSI situation. In the very limited S range of the data in the two
other situations the NLO prediction is significantly above, whereas the N$^2$LO
prediction is pretty close to the data, like in the case of CD-Bonn. Note that in
the  order N$^2$LO the EFT 3N forces are not yet included, which therefore is a 
preliminary and inconsistent result. In the case of N$^3$LO, again without the EFT 3N forces,
the bandwidth is somewhat larger than for N$^2$LO and there is a slight tendency for the
predictions to lie above the N$^2$LO ones. Since in both cases, N$^2$LO and N$^3$LO,
there is underbinding for $^3$H, one can possibly  expect, like for the conventional forces,
that the inclusion of the proper EFT 3N forces will shift the predictions downwards.
The underbinding at N$^3$LO is stronger than for N$^2$LO, which might compensate
the slightly higher values in N$^3$LO compared to N$^2$LO.

 Substantial differences between different orders show up in regions of the S curve
which have not been accessible to this experiment. A quantitative comparison reveals that
the conventional  force predictions lie close to the lower border of the N$^3$LO predictions in the minima.
 
\section{Summary}
We present new deuteron breakup cross-section data for the np 
FSI configuration, a coplanar star and an intermediate-star configuration.
The data are compared to predictions from nd breakup Faddeev calculations using 
the CD Bonn potential with and without the TM99' three-nucleon force,
 as well as with predictions from EFT which, however, in the case of N$^2$LO and N$^3$LO,
are still preliminary, since 3N forces at these orders required for consistency reasons
are not yet included.

Comparison with earlier results suggests that the good agreement between the 
data and theoretical predictions for the FSI situation follows the known 
systematics in this energy region and therefore should also apply for 
the analogue neutron-deuteron breakup. There is good agreement in the 
coplanar-star situation, but this may be fortuitous in view of the missing 
Coulomb interaction in the theoretical prediction. The slight overestimate 
of the data
 by the theory in the intermediate-star situation reminds of the space-star 
situation studied earlier, see e.g. \cite{pae01}. The EFT predictions are in 
excellent agreement with the data only for the FSI configuration. In the case of the CST
and IST situations  the N$^2$LO predictions are  rather close to the ones of CD-Bonn, but
that result has to be regarded as preliminary since the 3N force contributions
required for consistency reasons are still missing. This is even more the case for
N$^3$LO since there the $^3$H underbinding without proper 3N forces is larger than
for N$^2$LO. Calculations including those proper 3N forces are underway.

\newpage
\section*{Acknowledgements}
This work was supported by the Polish Committee for Scientific Research 
under grant no. 2P03B00825, and by DOE under grants nos. DE-FG03-00ER41132 and 
DE-FC02-01ER41187. This work was also supported by the U.S. Department of Energy Contract 
No. DE-AC05-84ER40150 under which the Southeastern Universities Research 
Association (SURA) operates the Thomas Jefferson Accelerator Facility. 
The numerical calculations have been performed 
on the Cray SV1 and T3E of the NIC in J\"ulich, Germany.


\newpage
\pagestyle{empty}

\begin{table}[htb]
\begin{center}
\begin{tabular}{l|ccc}
& CST& IST&FSI\\ \hline
$\Theta_3=\Theta_4$& 71.2&68.6&51.5 \\
$\Phi_3$&0 &15 &0 \\
$\Phi_4 $&180  &165   &180  \\
\end{tabular}
\end{center}
\caption{\label{winkel-tab} Laboratory angles of coincidence 
detectors for the three configurations.}
\end{table}
\newpage

\newpage
\newpage

\newpage

\newpage

\newpage

\newpage

\newpage


\newpage

\begin{figure}[htb]

\begin{center}
\epsfig{file=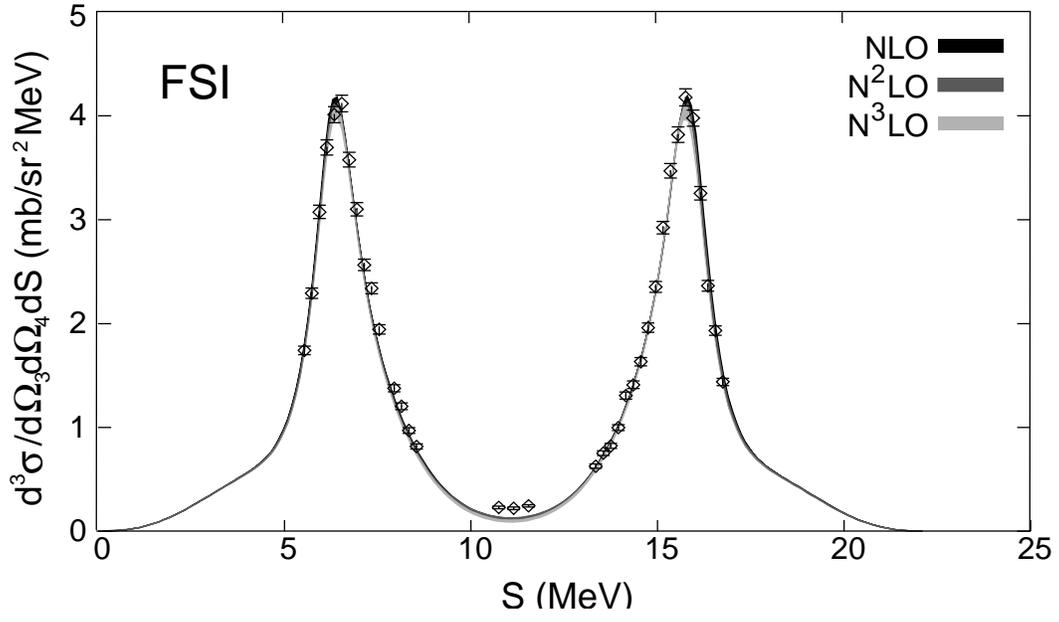,width=14cm}

\vspace{1cm}

\caption{\label{fsi-eft} Results of the breakup cross section
  $\frac{d^{3} \sigma}{ d \Omega_{3} d \Omega_{4} d S}$ 
(see table I) in the FSI situation. Experimental data are
  compared with the results of calculations in the framework ofthe EFT 
without a three-body force.}
\end{center}

\end{figure}
\newpage

\begin{figure}[htb]
\begin{center}

\epsfig{file=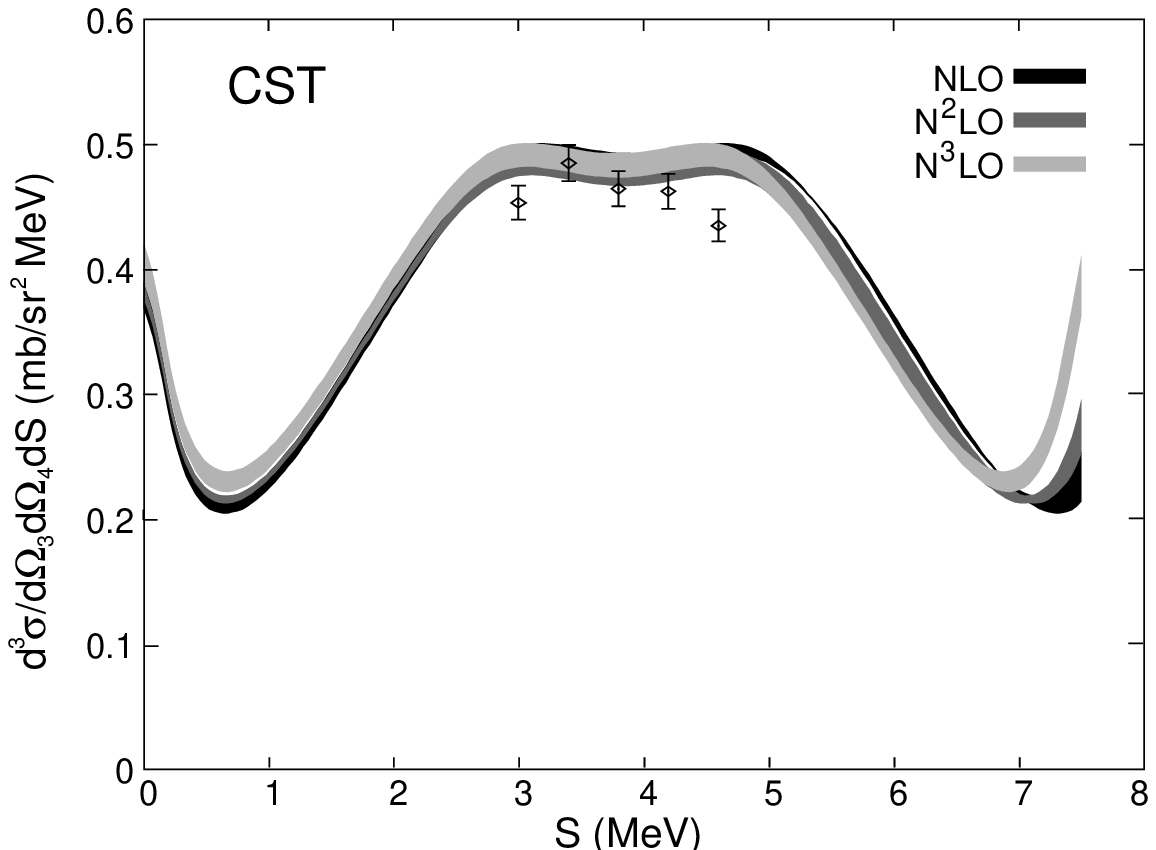,width=14cm}

\vspace{1cm}

\caption{\label{cst-eft} Results of the breakup cross section
  $\frac{d^{3} \sigma}{ d \Omega_{3} d \Omega_{4} d S}$ 
(see table I) in the CST  situation, similar to Fig. \ref{fsi-eft}.}

\end{center}
\end{figure}
\newpage
\begin{figure}[htb]
\begin{center}

\epsfig{file=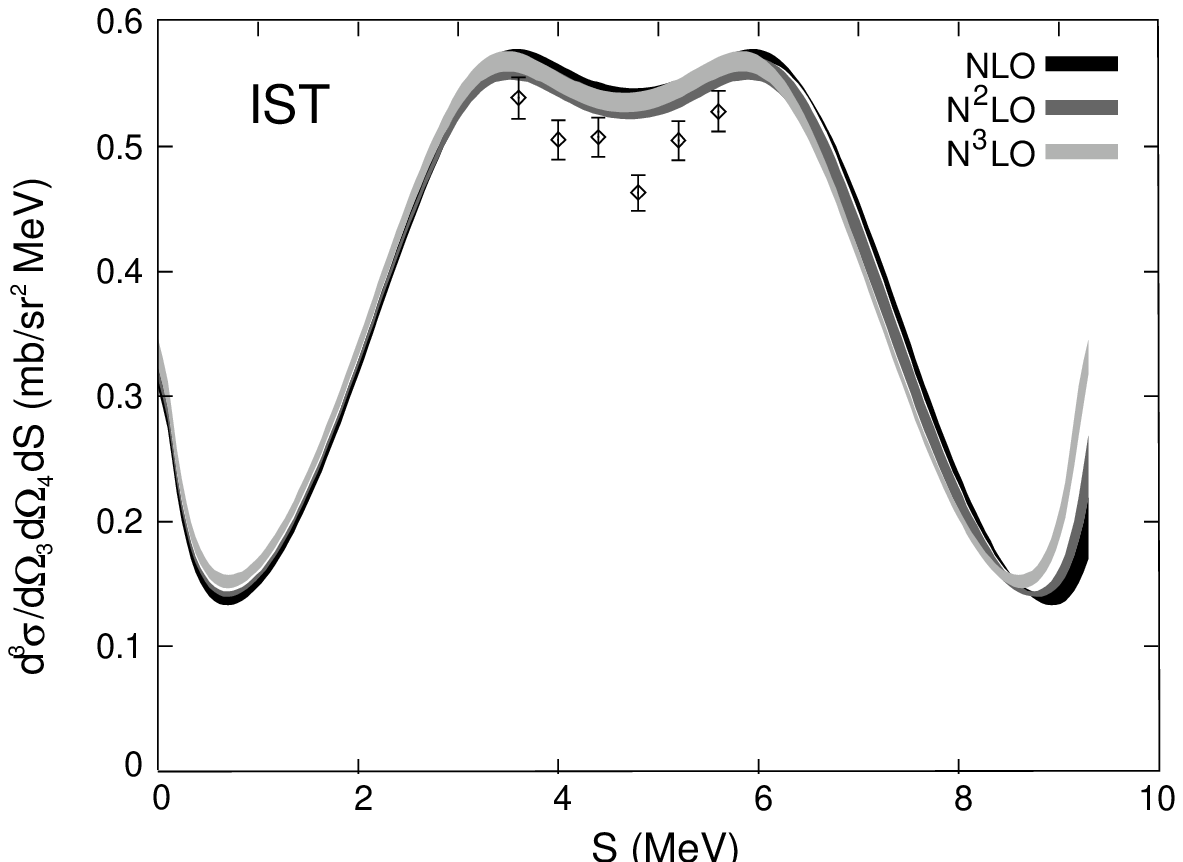,width=14cm}

\vspace{1cm}

\caption{\label{ist-eft} Results of the breakup cross section
  $\frac{d^{3} \sigma}{ d \Omega_{3} d \Omega_{4} d S}$ 
(see table I) in the IST  situation, similar to Fig. \ref{fsi-eft}.}
\end{center}
\end{figure}

\end{document}